# Strongly interacting Hofstadter states in magic-angle twisted bilayer graphene

Minhao He[1,2,†,#], Xiaoyu Wang[3,†], Jiaqi Cai[1], Jonah Herzog-Arbeitman[2], Takashi Taniguchi[4], Kenji Watanabe[5], Ady Stern[6], B. Andrei Bernevig[2], Matthew Yankowitz[1,7,#], Oskar Vafek[3,8,#], Xiaodong Xu[1,7,#]

[1]Department of Physics, University of Washington, Seattle, Washington, 98195, USA
[2]Department of Physics, Princeton University, Princeton, New Jersey 08544, USA
[3]National High Magnetic Field Lab, Tallahassee, Florida 32310, USA
[4] Research Center for Materials Nanoarchitectonics, National Institute for Materials Science, 1-1 Namiki, Tsukuba 305-0044, Japan
[5] Research Center for Electronic and Optical Materials, National Institute for Materials Science, 1-1 Namiki, Tsukuba 305-0044, Japan
[6]Department of Condensed Matter Physics, Weizmann Institute of Science, Rehovot 76100, Israel
[7]Department of Materials Science and Engineering, University of Washington, Seattle, Washington, 98195, USA
[8]Department of Physics, Florida State University, Tallahassee, Florida 32306, USA

† These authors contributed equally to the work.
#Correspondence to: minhaohe@princeton.edu, myank@uw.edu, vafek@magnet.fsu.edu, xuxd@uw.edu

**Abstract:** Magic-angle twisted bilayer graphene (MATBG) hosts a multitude of strongly correlated states at partial fillings of its flat bands[1–7]. In a magnetic field, these flat bands further evolve into a unique Hofstadter spectrum[8] renormalized by strong Coulomb interactions[9]. Here, we study the interacting Hofstadter states spontaneously formed within the topological magnetic subbands of an ultraclean MATBG device, notably including symmetry-broken Chern insulator (SBCI) states and fractional quantum Hall (FQH) states. The observed SBCI states form a cascade with their Chern numbers mimicking the main sequence correlated Chern insulators[10–15]. The FQH states in MATBG form in Jain sequence; however, they disappear at high magnetic field, distinct from conventional FQH states which strengthen with increasing magnetic field. We reveal a unique magnetic field-driven phase transition from composite fermion phases to a dissipative Fermi liquid. Our theoretical analysis of the magnetic subbands hosting FQH states predicts non uniform quantum geometric properties far from the lowest Landau level. This points towards a more natural interpretation of these FQH states as in-field fractional Chern insulators of the magnetic subbands.

## Introduction

The Hofstadter butterfly[8] is a general term describing the recursive energy band spectrum for electrons confined to two dimensions and with a magnetic length $l_B$ comparable to the lattice periodicity. A new class of topologically ordered states, termed in-field Chern insulators, emerges upon fully filling a set of the magnetic subbands within the Hofstadter butterfly[16]. Each such gapped state is characterized by an integer total Chern number, $t$, and integer moiré band filling factor, $s$, associated with its limit following the gap to zero magnetic field. Strongly correlated states, beyond those predicted within the non-interacting Hofstadter butterfly, can also emerge when Coulomb interactions become comparable to or exceed the width of the magnetic subbands. These interacting Hofstadter states arise upon partially filling the magnetic subbands and can be categorized into two distinct classes: those with fractional $s$ and integer $t$, and those with fractional $t$ and non-zero $s$. The former generally correspond to symmetry-broken Chern insulator (SBCI) states that enlarge the unit cell by spontaneously breaking the discrete translational symmetry of the moiré lattice[17,18]. The latter states instead result from electrons fractionalized into anyons and correspond to in-field fractional Chern insulator (FCI) states of the magnetic subbands[18]. Interactions could also generate states that exhibit the coexistence of charge fractionalization and a SBCI[18], necessarily arising at both fractional $s$ and $t$.

SBCI states and in-field FCI states have been seen previously in moiré superlattices of monolayer or Bernal bilayer graphene aligned with hexagonal boron nitride[17,18] (hBN). However, these moiré lattices require extremely high magnetic fields of $B \gtrsim 20$ T to generate such correlated phases owing to their large electronic band dispersions at zero field. In contrast, the flat electronic bands of MATBG (and other twisted graphene multilayers) greatly enhance the strength of Coulomb interactions and promote the formation of correlated states at much lower $B$[10,19–22], making them ideal platforms to study the properties of strongly interacting Hofstadter states. Our study focuses on MATBG as a representative flat-band moiré system. Fingerprints of strong interactions at finite magnetic field have been previously seen in the form of a characteristic sequence of low-field correlated Chern insulators[10–15] (CCI). Regardless of slight differences in twist angle and uncontrolled variations in the microscopic strain distribution, twisted bilayer graphene devices near the magic angle always exhibit similar CCI having Chern numbers $|t| = 1, 2, 3$ with corresponding moiré filling factors $|s| = 3, 2, 1$, respectively. Theoretically reproducing this robust experimental feature has only recently been achieved in an interacting Hofstadter calculation[9] that includes uniaxial heterostrain. Experimentally, there has been only limited experimental attention[4,10,19,20] focused on understanding the correlated Hofstadter states with fractional $t$ or $s$ arising within the interaction-renormalized magnetic subbands.

By studying an ultra-high quality MATBG device, we report the observation of a cascade of SBCI states that forms primarily upon hole doping its magnetic subbands. The sequence of Chern numbers of these states is closely connected to an analogous sequence for the parent CCI states, in that both form a cascade where the Chern number $|t|$ changes in increments of 1. These sequences are well captured by a self-consistent Hartree-Fock calculation of the interacting Bistritzer-MacDonald (BM) model in a finite magnetic field. We additionally see a sequence of fractional quantum Hall (FQH) states (fractional $t$ and $s = 0$) formed within the magnetic subbands emanating from the charge neutrality point (CNP). Although these states follow the usual Jain sequence[23], their appearance at the field regime where the magnetic length is comparable to the moiré periodicity poses a key distinction from FQH states seen in typical Landau levels, pointing to their unconventional nature. This is experimentally signified by the phase transitions from FQH states into a normal Fermi liquid as the magnetic field is raised above ≈10 T. This behavior falls outside the usual theoretical paradigm for understanding FQH states, and is in strong contrast to the usual monotonically-enhanced FQH states in crystalline graphene and most other two-dimensional electronic gases. Our HF band analysis shows that these fractional states arise out of strained magnetic subbands with finite bandwidth and non-uniform, non-ideal quantum geometric properties. Unlike in pristine LLs, this unusual quantum geometric structure points to a potential description of these states within the framework of magnetic-field–induced FCIs.

**High quality MATBG/WSe₂ sample and its Landau fan diagram**
We focus our attention on a MATBG device with a twist angle of $\theta = 1.03°$ (Fig. 1a-b). The presence of monolayer WSe$_2$ has been argued to stabilize the twist angle and reduce the resulting moiré disorder[24]. This can be seen from a reflection spectroscopy study on the same device using the exciton Rydberg state of monolayer WSe$_2$ as a local optical sensor, which shows that the sample is highly uniform[25]. The WSe$_2$ also induces spin-orbit coupling in MATBG[26,27] with an energy scale of $\approx$1 meV; although this may in principle change the correlated grounds states, most of our results likely also generalize to MATBG without WSe$_2$ due to the overall similar phenomenology previously observed in such devices. Figure 1c shows the longitudinal resistivity, $\rho_{xx}$, as a function of moiré filling factor, $\nu$, measured at a temperature of $T = 340$ mK (see Fig. S1 for temperature dependence). In addition to the correlated insulating states at moiré band filling factors of $\nu = -2$, +2, +3, we observe superconductivity (SC) at $\nu = -2 - \delta$ (Fig. 1d), and orbital ferromagnetism[28] associated with anomalous Hall effects (AHE) near $\nu = +1$ and +2 (Fig. 1e). Both instances of the AHE show typical hysteresis loops in the Hall resistance, $\rho_{yx}$, with amplitude of $\approx$1 kΩ (see Fig. S2 for the dependence on $\nu$). Overall, our observations are consistent with previous studies of MATBG both with[24,26,27] and without[29–31] a WSe$_2$ substrate.

The simultaneous observation of SC and AHE in a single device is rare[24,29], and partially signifies the high sample quality. The unprecedented homogeneity of our device is best reflected in the rich Landau fan diagrams of both $\rho_{xx}$ and $\rho_{yx}$ (Fig. 1f, Fig. S3), which will be the focus of our study. A comprehensive phase diagram at both zero and finite magnetic field is shown in Fig. 1g. The Hofstadter states at finite magnetic field are identified by the simultaneous suppression of $\rho_{xx}$ and quantization of $\rho_{yx}$ to values of $h/te^2$, corresponding to their Chern number $t$. States without fully quantized $\rho_{yx}$ are identified by the slope of their trajectories defined by the Streda formula[32], $t = (h/e)(\partial n/\partial B)$. Integer quantum Hall states and in-field Chern insulators having both integer $t$ and $s$ form the primary features of the Landau fan at finite magnetic field, and are represented by purple lines in the schematic (see also the Landau fan at 2 K in Fig. S4). Close inspection of the fan diagram further reveals strongly correlated states beyond the single particle Hofstadter spectrum, including numerous SBCI states (red lines) and FQH states (yellow lines). We will elaborate on the nature of these states in the remainder of our discussion.

**Cascades of symmetry-broken Chern insulators**
Figures 2a,c show high-resolution zoom-ins on two groups of SBCI states observed on the hole-doped side of the Landau fan (see $\rho_{yx}$ in Fig. S5), with the most robust gapped states denoted schematically in Figs. 2b,d, respectively. In addition to the conventional hierarchy of CCIs reported across many devices previously, we see an associated cascade of SBCI states indicated by the red lines in Fig. 2b. Notably, these states exhibit the same cascading sequence of Chern numbers ($t = -3, -2, -1$) as their adjacent parent CCI, but with half-integer moiré filling indices ($s = -1/2, -3/2, -5/2$). We adopt the notation ($t, s$) for simplicity in referencing each of these states. Figure 2e shows linecuts of $\rho_{xx}$ and the Hall conductivity, $\sigma_{yx}$, calculated from tensor relation $\sigma_{yx} = \rho_{yx}/(\rho_{yx}^2 + \rho_{xx}^2)$ at $B = 8.5$ T. The (–2, –3/2) and (–3, –1/2) SBCI states are almost perfectly developed, with vanishing $\rho_{xx}$ and nearly-quantized $\sigma_{yx}$, whereas the (–1, –5/2) SBCI is clearly visible in the

Landau fan but is not fully quantized. We determine the associated energy gaps, Δ (Fig. 2c inset, see also Fig. S6), from the thermal activation behavior of $\rho_{xx} \propto e^{\Delta/2k_BT}$, where we find that the (–2, –3/2) and (–3, –1/2) states have energy gaps about three times larger than the (–1, –5/2) state. The results from the former two states are consistent with previous studies[10,20], whereas the latter has not yet been seen until our work. Together, these three SBCI states form a new cascade with an equal interval of $\Delta t = -1$ and $\Delta s = 1$ between each state, exactly as describes the main sequence of CCIs but shifted by $s_0 = 1/2$. These SBCI states can be understood as maintaining the same spin and valley polarization properties of their parent CCI[10], but further breaking the moiré lattice translational symmetry by forming a charge density wave that spontaneously doubles the unit cell area.[10]

Intriguingly, we find that these three SBCI states disappear at the same lower threshold magnetic field, corresponding to a simple rational flux ratio $\Phi/\Phi_0 = 1/4$. A second group of SBCI states, (–3, –2/3) and (–4, –1/3), are also found to be closely tied with a simple rational flux ratio $\Phi/\Phi_0 = 1/3$ (Fig. 2c-d and f, see also Fig. S6). These two SBCI states are flanked by the (–2, –1) and (–5, 0) Chern insulators. Together, these bound a magnetic subband with Chern number C = –3 emanating from $\Phi/\Phi_0 = 1/3$ towards zero field (green shaded region). The two SBCI states can be intuitively understood as partitioning the C = –3 subband into three parts, each with C = –1.

Although a similar cascade of SBCI states is naively expected on the electron doping side in the presence of approximate particle-hole symmetry, we do not see this in our Landau fan. At high magnetic field, we only observe a weak (1, 5/2) SBCI state within this sequence (Fig. S7). The (2, 3/2) and (3, 1/2) states were previously reported in a different sample[10], but are absent here. However, the (2, 3/2) SBCI is observed over a small range of magnetic field between ≈2–3 T nearby the correlated insulator at ν = 2. It is currently not clear why the electron-doped side of our sample deviates from the behavior of the hole-doped side, as well as from prior samples and theoretical expectation.

**Hartree-Fock calculations at finite magnetic field**
We obtain more insights on the observed SBCI states by turning to in-field self-consistent Hartree-Fock calculations[9]. Here, we perform HF calculations of the interacting Bistritzer-MacDonald (BM) Hamiltonian at twist angle θ = 1.05° and crucially include uniaxial heterostrain $\epsilon$ = 0.2%. More details of the model calculation can be found in SI. Figure 3a shows the calculated phase diagram of the incompressible states formed between magnetic flux $\Phi/\Phi_0 = 1/12$ and 1/2, with the marker radius denoting the size of the associated charge gap. We further characterize each data point by examining the wavefunction, isospin symmetry, and moiré translation symmetry (Fig. S8), and group them according to their topological indices (t, s). States with the largest energy gaps form the main sequence of the CCI states with |t| = 1, 2, 3 and |s| = 3, 2, 1, respectively, as well as the integer quantum Hall states with |t| = 4, 3, 2, 1 and |s| = 0, and are denoted using purple. The black markers represent the topologically trivial correlated insulators at |s| = 2, 3, which corresponds to the states marked by the black lines in Fig. 1g. These t = 0 states are adiabatically connected to the zero-field ground state (marked by magenta in Fig. 1g), which are now commonly believed to be incommensurate Kekulé spiral (IKS) states in presence of strain[33,34]. A detailed theoretical analysis of the |s| = 3, 2 states also reveals their IKS like isospin symmetry breaking (Fig. S9).

The calculation further predicts a series of Chern insulators with $|t|$ = 1, 2, 3 and $|s|$ = 5/2, 3/2, 1/2 respectively, denoted in red, along with a (-3, -2/3) state with smaller energy gap. All correspond to SBCI states seen in our Landau fan. Figure 3b shows the calculated energy spectrum of the (-3, -1/2) SBCI gap as function of magnetic flux (see more in Fig. S10). The interaction-driven gap size changes non-monotonically with magnetic field, with an apparent energy gap opening for $\Phi/\Phi_0 > 1/8$ and closing for $\Phi/\Phi_0 \approx 1/2$. The calculated local density of states (LDOS) at the edge of the gap exhibits a real-space charge modulation forming a stripe like feature (Fig. 3c and Fig. S11). The fractional filling factor $s$, nonzero Chern number $|t|$, and the broken moiré translation symmetry are all consistent with our interpretation of these states as SBCIs.

Overall, we find a good qualitative agreement between theoretical calculations and the experimental phase diagram of our ultraclean MATBG sample. This can be further demonstrated by comparing the calculated and measured energy gaps of the three half-integer SBCIs (Fig. 3a inset versus Fig. 2e inset). Although the experimentally measured transport gaps are likely underestimated due to the presence of (twist angle) disorder[35], we find they follow the same relative hierarchy as in the theorical prediction. Lastly, we note that the calculation captures an upper threshold for the (-3, -2/3) state at $\Phi/\Phi_0 = 1/3$, the same as the experimental result. Yet it does not capture the experimentally determined lower threshold at $\Phi/\Phi_0 = 1/4$ of the half-integer SBCI states. This threshold appears to arise due to a competition with other Chern insulator states denoted by gray lines in Fig. 2b, and will require additional modeling work to understand in detail. Capturing the distinct behavior on the electron-doped side will also likely require further corrections to the BM model. Interestingly, our HF calculations also predict intervalley coherence underlying the SBCI states (Fig. S12), much like the parent CCI states. Future microscopic studies using scanning tunneling microscopy will be critical to reveal the broken moiré translation symmetry and explore the potential Kekulé pattern[34] of the SBCI states.

**Unconventional fractional quantum Hall states in magnetic subbands**
We next discuss the observation of fractional quantum Hall states in our sample, and highlight their unconventional nature. Figure 4a shows a map of $\rho_{xx}$ taken at high magnetic field, in which we observe FQH states at certain partial fillings of each magnetic subband with effective filling factors $\widetilde{v_c} = v_c - [v_c]$ = 1/3, 2/5, 3/5, 2/3, where $v_c = n\Phi_0/B$ is the Landau level filling fraction. These fillings are consistent with the usual Jain sequence[23], as is generally observed in the lowest LL with orbital number $N$ = 0. Figure 4b shows linecuts of $\sigma_{yx}$ and the longitudinal conductivity, $\sigma_{xx} = \rho_{xx}/(\rho_{yx}^2 + \rho_{xx}^2)$, at $B$ = 7 T, exhibiting nearly quantized $\sigma_{yx} = \frac{v_c e^2}{h}$ at $v_c$ = -1/3, -2/3, -4/3, -5/3, -7/3 and -8/3. The values of $\sigma_{yx}$ at $v_c$ = -10/3, -11/3 deviate more substantially from their anticipated values due to mixing with the larger residual $\sigma_{xx}$. The inset of Fig. 4b shows measurements of the thermal activation gaps of the FQH states with denominator 3 (see also Fig. S13). All of the gaps are close to 1 K, and in certain cases do not respect particle-hole symmetry relative to half filling. These energy gaps are one order of magnitude smaller compared to Laughlin states at $v_c$ = 1/3, 2/3 in monolayer graphene[36–40] and bilayer graphene[41–45]. They are about $0.01E_c$, assuming a Coulomb energy $E_c = e^2/4\pi\epsilon\epsilon_0 l_B$ with effective dielectric constant $\epsilon \approx 15$. Besides the inevitable moiré disorder[35] which is absent in crystalline graphene, the small energy gaps of the FQHs could also be weakened by competing Fermi liquid phase due to the finite bandwidth of magnetic subbands, or by charge/spin density wave instabilities due to strain and non-uniform quantum geometry.

Remarkably, the FQH states disappear at high magnetic fields (above ≈8.5-10 T). This is distinct from the usual case of LLs, where the FQH gaps scale with the strength of Coulomb interactions and thus monotonically increase with magnetic field. Fig. 4c shows the linecuts of $\rho_{xx}$ and $\rho_{yx}$ as a function of $B$, showing the evolution of FQH states at $\nu_c$ = -4/3, -5/3 (top panel) into a gapless Fermi liquid. We see similar crossover behavior at half-filling, $\nu_c$ = -3/2 (bottom panel), which is a putative composite Fermi liquid when flanked by FQH states. Both panels show a smooth phase transition from the Jain sequence FQH states to a dissipative Fermi liquid phase as $B$ increases (see Fig. S14). Intuitively, this can be understood as the FQH states becoming destabilized by the broadened bandwidth at a higher magnetic field, and eventually losing the competition to a normal Fermi liquid.

Similar non-monotonic behavior of FQH states was previously observed in aligned graphene/h-BN[17,18], yet little is known about the nature of these states. In moiré lattices, FQH states appear at partial filling of the interaction-renormalized magnetic subbands emanating from CNP. For MATBG these are commonly referred to as the zero energy LLs, but despite their name, the corresponding energy bands are in fact distinct from Landau levels due to the non-uniform Berry curvature distribution. Rather, they are strongly influenced by the moiré pattern, and should be thought of as magnetic subbands within the Hofstadter butterfly. They feature finite dispersion and arise by hybridizing different low-energy LLs from the CNP of the two graphene layers. As a result, they may differ significantly from the physics of the $N$ = 0 lowest Landau level[46] even when they have the same Chern number, as we elaborate upon below.

Figure 4d shows the interaction-renormalized energy spectrum when the Fermi energy is tuned into the gap of the (-4, 0) state, with red color denoting the magnetic subbands hosting the observed FQH states. Although a detailed theoretical modeling of the FQHs by exact diagonalization is beyond the scope of our work, we qualitatively evaluate their nature by inspecting the quantum geometry of the relevant magnetic subbands. We adopt the notion of ideal quantum geometry indicators[47] which evaluate the similarity of the subband with the $N$ = 0 LL (see SI for details), and calculate whether the magnetic subbands are close to the ideal conditions needed for hosting fractionalized states. Figure 4e-f show the distribution of the imaginary part of the quantum geometry $\eta$, i.e. Berry curvature, $\mathcal{F}$, and the real part of $\eta$, i.e. quantum metric, $g$, of the red magnetic subbands. Notably, both $\mathcal{F}$ and $g$ exhibit a non-uniform profile within the magnetic Brillouin zone, distinct from the $N$ = 0 LL with uniform distribution of the value $2\pi$. The uniformity of the Berry curvature is measured by its standard deviation, $\sigma(\mathcal{F}) \approx 0.113$, which is finite unlike the $N$ = 0 LL but nevertheless relatively small. The trace condition measures the similarity of the subband's wavefunctions to the $N$ = 0 LL, and is found to be $T(\eta) \approx 4.226$ (see SI for detail). In contrast to the relatively uniform Berry curvature distribution, the large trace condition indicates dissimilarity with the lowest LL (where $T(\eta) = 0$).

**Conclusion**
Nearly uniform Berry curvature is known to support fractionalized states, allowing us to rationalize the appearance of the Jain sequence FQH states even in bands quite unlike the lowest Landau level. This is further confirmed by the decomposition of the magnetic subbands into the basis of the Landau levels of monolayer graphene (Fig. S15), from which we find the majority contribution comes from LLs with $N \neq 0$. Despite their emanation from the CNP like conventional FQH states, our calculations suggest the fractionalized states we observe are closer to in-field FCIs arising out of a lattice Chern band at zero magnetic field. These states require a magnetic field to form and arise despite the quantum geometric non-idealities of the magnetic subbands, including their finite bandwidth, quantum metric fluctuations, and the small gaps to nearby bands. Similar arguments should be applicable to an entire class of FQH states in moiré superlattices where the quantum geometry is non-uniform and the magnetic length is comparable to the superlattice period.

## Methods

### Sample fabrication
Heterostructures of graphite/hBN/WSe$_2$/MATBG/hBN/graphite are assembled using a standard dry-transfer technique with a PC/PDMS (polycarbonate/polydimethylsiloxane) stamp and transferred onto a Si/SiO$_2$ wafer. MATBG stack is fabricated by using the tear-and-stack method. We use 3-5 nm graphite as bottom gates whereas 3-5 layers graphene are used for top gate to minimize its optical absorption. We have no intentional control over the twist angle between WSe2 and its adjacent graphene layer during the stacking process. CHF$_3$/O$_2$ and O$_2$ plasma etching followed by electron beam lithography are used to define a Hall bar geometry, and Cr/Au contacts (7nm/70nm) are finally added using electron beam evaporation.

### Transport measurements
Transport measurements are conducted in a Bluefors dilution refrigerator. Measurements are performed with a 1-5 nA a.c. excitation current at either 13.3 Hz or 13.7 Hz. The current and voltage are pre-amplified by DL 1211 and SR560 respectively, and then read out by SR830/SR860 lock-in amplifiers. Gate voltages are supplied by either NI DAQ or Keithley 2450. In this study, we use bottom gate voltage $V_b$ to control carrier density n: $n = V_b C_b/e$, where $C_b$ is the bottom gate capacitances per unit area. A global Si gate voltage is also applied to reduce the contact resistance.

The filling factor, $\nu$, is defined as the number of electrons per moiré unit cell. Full filling of the eight-fold degenerate flat bands in TBG corresponds to 4 electrons or holes per moiré unit cell, $\nu= \pm 4$. The carrier densities corresponding to the full fillings $\nu= \pm 4$ can be determined by measuring the Hall coefficient at a small magnetic field $B$, $R_H = [\rho_{xy}(B) - \rho_{xy}(-B)] / (2B)$. The Hall carrier density, $n_H = 1/eR_H$, reverses carrier type (from electron to hole, or vice versa) across the full fillings $\nu= \pm 4$ and can be identified as the zero crossing in the $R_H$ measurements. The twist angle is then determined following the relationship $n = 8\theta^2/\sqrt{3}a^2$, where $a = 0.246$ nm is the graphene lattice constant. The twist angle is further confirmed by fitting the observed quantum Hall states and Chern insulators within the allowed Hofstadter states in the Wannier diagram.

## Acknowledgements
The work at UW is supported by NSF MRSEC DMR-2308979. M.H. acknowledges the support from the Princeton quantum initiative. X.W. acknowledges the support from the National High Magnetic Field Laboratory through NSF Grant No. DMR-2128556 and the State of Florida. BAB was supported by Simons Investigator Grant No. 404513. BAB and OV are supported by the Gordon and Betty Moore Foundation's EPiQS Initiative (Grant No. GBMF11070). K.W. and T.T.

acknowledge support from the JSPS KAKENHI (Grant Numbers 21H05233 and 23H02052) and World Premier International Research Center Initiative (WPI), MEXT, Japan.

**Author Contributions**
M.H. fabricated the device. M.H. and J.C. performed the measurements, analyzed the data, supervised by M.Y. and X.X. X.W. performed the theoretical calculations, supervised by O.V., also with theory inputs from J.A.H, A.S. and B.A.B. K.W. and T.T. grew the BN crystals. M.H., X.W., M. Y., O.V. and X.X. wrote the paper with input from all authors.

**Corresponding author**
Correspondence to Minhao He, Matthew Yankowitz, Oskar Vafek or Xiaodong Xu.

**Data Availability**: Source data are provided with this paper. All other data are available from the corresponding author upon reasonable request.

**Code Availability:** Codes used for data analysis in this study are also available from the corresponding author upon reasonable request.

**Competing Interests**
The authors declare no competing interests.

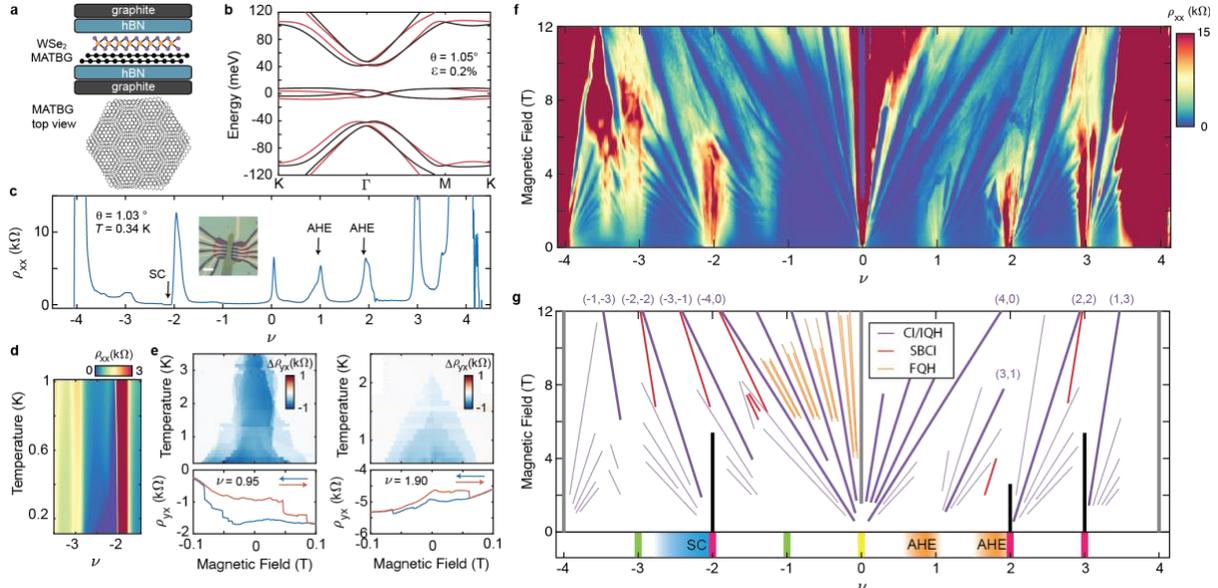

**Figure 1. Transport characterization of a $\theta = 1.03°$ high quality MATBG/WSe$_2$ sample. a,** Schematic of the device structure. **b,** Flat bands of the MATBG calculated at twist angle of $\theta = 1.05°$, and uniaxial heterostrain $\epsilon = 0.2\%$ and $\varphi = 0°$ (see SI for details). **c,** $\rho_{xx}$ versus $\nu$ measured at $T = 340$ mK. Regions of observed superconductivity and anomalous Hall effects are denoted by arrows. The insets shows an optical micrograph of the sample. The scale bar is 5 μm. **d,** Temperature dependence of $\rho_{xx}$ in the vicinity of superconductivity. **e,** Measurements of $\rho_{yx}$ as $B$ is swept back and forth. Anomalous Hall effects near $\nu = 1$ (left panels) and $\nu = 2$ (right panels). The top panels show the amplitude of the hysteresis loop, $\Delta\rho_{yx} = \rho_{yx}^{B\downarrow} - \rho_{yx}^{B\uparrow}$, acquired as function of temperature. The bottom panels show line traces measured at $T = 300$ mK. **f,** Landau fan diagram of $\rho_{xx}$ measured up to 12 T at $T = 300$ mK (the corresponding $\rho_{yx}$ is shown in Fig. S3). **g,** Schematic diagram indicating the different phases seen in the devices. At zero magnetic field, we denote superconductivity with the blue rectangle and regions exhibiting the anomalous Hall effect with orange. Tentative assignments of the correlated states at integer fillings are also denoted, gapped IKS with magenta color, gapless IKS with green color, and semi-metallic phase with yellow color. At finite field, we denote Chern insulators and integer quantum Hall states with purple lines, SBCI states with red, FQH states with yellow, and the topologically trivial insulators states with black.

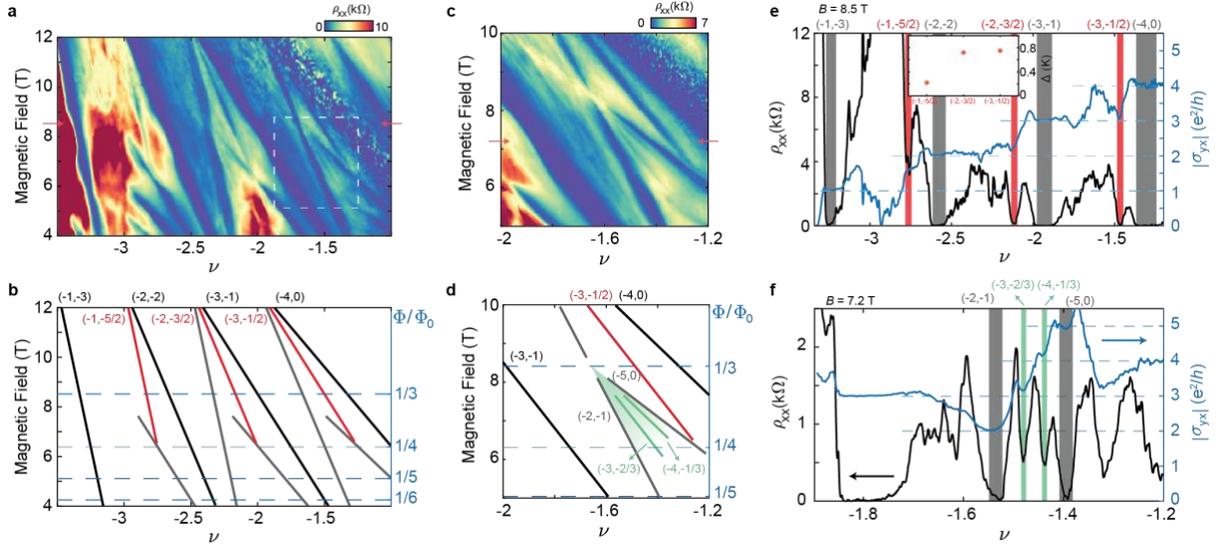

**Figure 2. Cascades of SBCI states. a,** Zoom-in measurement of the $\rho_{xx}$ Landau fan focusing on the SBCI states on the hole-doped side. **b,** Schematic diagram of the most robust gapped states. The left y-axis shows the corresponding magnetic flux ratio, $\Phi/\Phi_0$. **c,** Further zoom-in of the region denoted by the white dashed box in **a**. **d,** Schematic diagram of the most robust gapped states. The green shading indicates a magnetic subband with a total Chern number of -3. **e,** $|\sigma_{yx}|$ (right axis) and $\rho_{xx}$ (left axis) measured at $B = 8.5$ T (red arrows in **a**). Shaded regions mark the SBCI states (red) and the main-sequence CCIs (gray). The inset shows measurements of the thermal activation gap of the three SBCI states denoted in **b**. **f,** The same measurement at $B = 7.2$ T (red arrows in **d**). The green shading indicates the SBCI states seen in **d**.

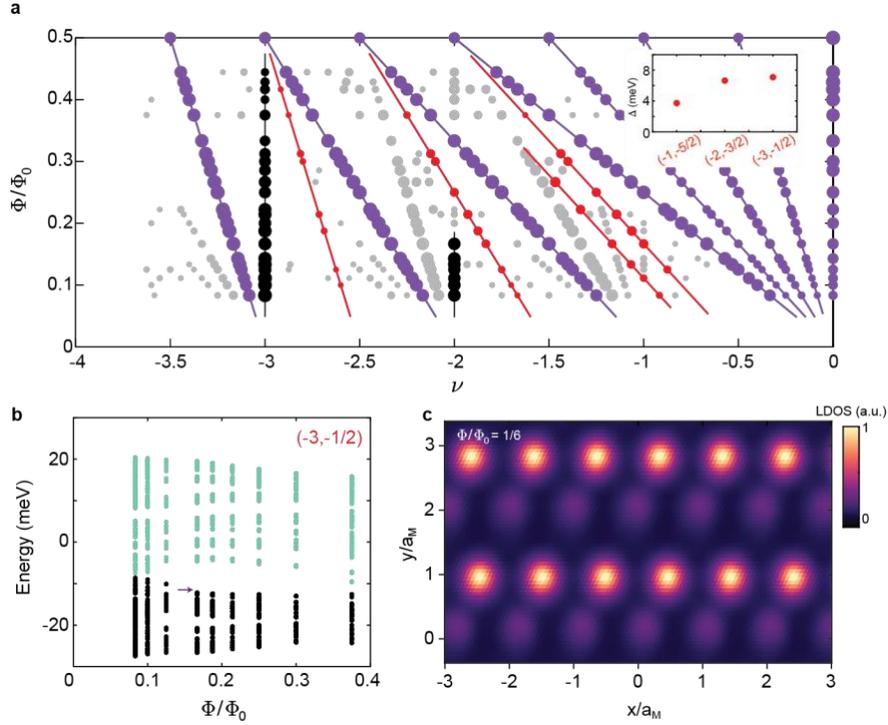

**Figure 3. Finite-field Hatree-Fock calculation of the SBCI states. a,** Calculated gapped states at finite magnetic field. The marker size is proportional to the gap size at each filling factor $\nu$ and magnetic flux ratio $\Phi/\Phi_0$. The gapped states originating from integer values of $\nu$ are denoted by purple and gray coloring. Purple states here match the states denoted in purple from Fig. 1g. The SBCI states are denoted by red coloring. Robust topologically trivial states are denoted by black coloring. The inset shows the calculated gap size of the three SBCIs at $\Phi/\Phi_0 = 3/10$. **b,** Interacting Hofstadter spectrum when the Fermi energy is in the gap of the (-3,-1/2) state. Occupied and unoccupied states are colored in black and green, respectively. **c,** LDOS calculated for the valence band edge of the (-3, -1/2) SBCI state at $\Phi/\Phi_0 = 1/6$ (purple arrow in **b**). $a_M \equiv \sqrt{|L_1||L_2|}$ is the effective periodicity of the strained moiré superlattice.

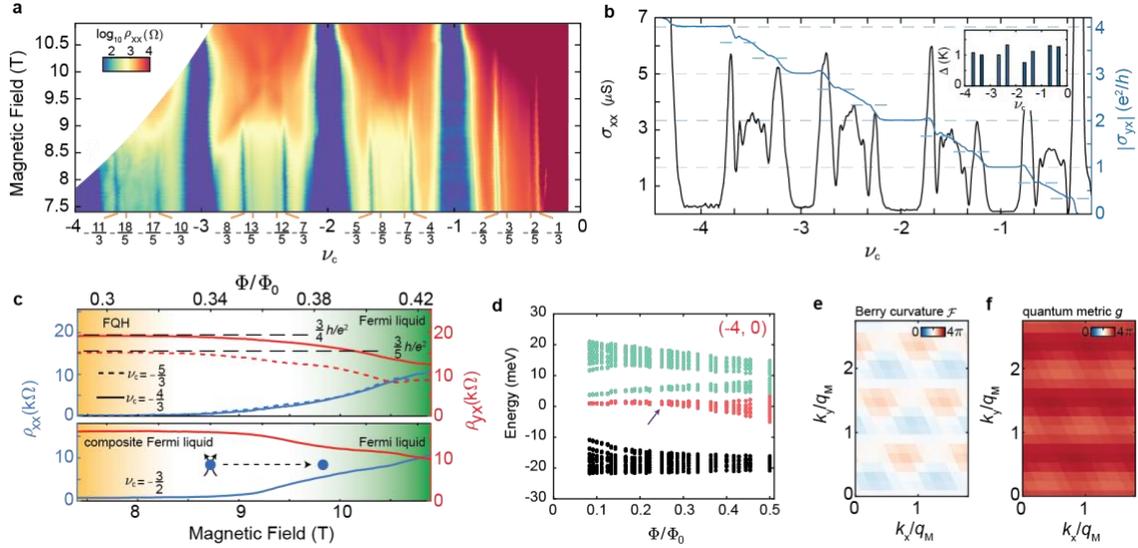

**Figure 4. Unconventional FQH states in magnetic subbands with finite bandwidth. a,** Zoom-in measurement of the $\rho_{xx}$ Landau fan, plotted against $\nu_c$. **b,** $\sigma_{xx}$ (left y-axis) and $\sigma_{yx}$ (right y-axis) measured at $B = 7$ T. The inset shows the measured thermal activation gaps of the FQH states with denominator 3. **c,** $\rho_{xx}$ and $\rho_{yx}$ measured as a function of $B$ at $\nu_c = -5/3$ and $-4/3$ (top panel) and $\nu_c = -3/2$ (bottom panel). The FQH states and composite Fermi liquid at low magnetic field transition to a normal Fermi liquid at high magnetic field. **d,** Calculated interacting Hofstadter spectrum when the Fermi energy is in the gap of the (-4, 0) state. Occupied states are marked by black, unoccupied states are marked by green and red. The red colored bands are the magnetic subband hosting the observed FQH states. **e, f,** Berry curvature $\mathcal{F}$ and quantum metric $g$ of the red magnetic subband, calculated at $\Phi/\Phi_0 = 1/4$ (purple arrow in **d**). $q_M$ is defined following $q_M \equiv \sqrt{|g_1||g_2|}/q$ at flux ratio $\Phi/\Phi_0 = p/q$.

# Supplementary information:

# Strongly interacting Hofstadter states in magic-angle twisted bilayer graphene

# Table of contents

We first briefly summarize the conventions used in this work for constructing the finite magnetic field phase diagram based on the interacting Bistritzer-MacDonald (BM) model in the presence of uniaxial heterostrain, the procedure for calculating the STM local density of states (LDOS), as well as the fractional Chern insulator (FCI) indicators of interaction-renormalized magnetic subbands. For more detailed discussions we refer interested readers to the Supplementary Information (SI) in Ref. [1].

**Conventions**

Uniaxial heterostrain refers to applying uniaxial tensile strain to one layer of graphene, and compressive strain of equal strength along the same axis to the other layer. It is parameterized by the heterostrain strength $\epsilon$ and orientation $\varphi$. We define the orientation with respect to the zigzag edge of untwisted and undeformed monolayer graphene. The impact of uniaxial heterostrain is amplified on the moiré length scale, causing deformations of the moiré (reciprocal) lattice vectors as well as modifying the energy dispersions of the moiré flat bands [2,3]. We hereby use the notations $L_{i=1,2}$ to denote the deformed moiré lattice vectors, and $g_{i=1,2}$ to denote the deformed reciprocal lattice vectors after the application of strain. They satisfy $L_i \cdot g_j = 2\pi \delta_{ij}$. We also use the convention that the top (bottom) graphene layer is twisted counterclockwise (clockwise) by $\theta/2$, and define the $+\hat{z}$ direction along the out-of-plane axis from the bottom layer to the top layer.

In a finite magnetic field along the $+\hat{z}$ direction $\mathbf{B} = B\hat{z}$, the BM model of a given valley ($\eta = K, K'$) and spin ($s = \uparrow, \downarrow$) flavor, $\hat{H}_{\eta,s}(\hat{\mathbf{p}})$, is modified to $\hat{H}_{\eta,s}(\hat{\mathbf{p}} + e\mathbf{A}(\mathbf{r}))$, where $\hat{\mathbf{p}} \equiv -i\hbar \boldsymbol{\nabla}$ is the momentum, $e > 0$ is the absolute charge of an electron, and $\mathbf{A}(\mathbf{r})$ is the magnetic vector potential satisfying $\boldsymbol{\nabla} \times \mathbf{A}(\mathbf{r}) = \mathbf{B}$. We shall choose the Landau gauge $\mathbf{A}(\mathbf{r}) \equiv Bx\hat{y}$ where the $\hat{y}$ is defined to be along the strain-deformed $\mathbf{L}_2$.

## Note S1. Self-consistent Hartree-Fock calculation of MATBG at finite field.

The finite magnetic field studies are performed at rational magnetic flux ratios $\phi/\phi_0 = p/q$ where $\{p,q\}$ are coprime integers, $\phi = \mathbf{B} \cdot (\mathbf{L}_1 \times \mathbf{L}_2)$ is the flux per moiré unit cell, and $\phi_0 = h/e$ is the magnetic flux quantum. At these flux ratios, the interacting BM model preserves the magnetic translation symmetry generated by

$$\hat{t}_{L_1}(r) \equiv e^{-i2\pi\frac{\phi}{\phi_0}\left(\frac{y}{|L_2|} - \frac{L_{1y}}{2|L_2|}\right)} \hat{T}_{L_1}, \quad \hat{t}_{L_2} \equiv \hat{T}_{L_2},$$

with $\hat{T}_{L_{i=1,2}}$ the discrete moiré translations at $B = 0$. The magnetic translation operators do not commute $[\hat{t}_{L_1}(r), \hat{t}_{L_2}] \neq 0$, however they satisfy $[\hat{t}_{L_1}(r), \hat{t}_{L_2}^q] = 0$. Therefore, to study the finite $B$ problem we enlarge the unit cell to be $L_1 \times qL_2$. Accordingly, we can define "magnetic Bloch states" labeled by the magnetic wavevector $k = k_1 g_1 + k_2 g_2$, where $k_1 \in [0,1)$ and $k_2 \in [0,1/q)$. The magnetic Bloch states $|\Psi_a^{(\eta s)}(k)\rangle$ are constructed from the monolayer Landau level wavefunctions, with more details presented in Ref. [1]. They are labeled by the magnetic subband index $a = 1, \ldots 2q$, and we order these states by increasing energy of the non-interacting BM model. In contrast to $B = 0$, here there are $2q$ magnetic subbands per valley and spin instead of 2. However, the magnetic Brillouin zone is also $q$ times smaller, therefore preserving the dimension of the Hilbert space between $B = 0$ and $B \neq 0$.

The interacting BM model projected onto the narrow moiré bands can be written as follows:

$$H = \sum_{\eta s \mathbf{k}} \varepsilon_a^{(\eta s)}(\mathbf{k}) d^\dagger_{\eta s a, \mathbf{k}} d_{\eta s a, \mathbf{k}} + \frac{1}{2A} \sum_{\mathbf{q}} V_{\mathbf{q}} \delta \hat{\rho}_{\mathbf{q}} \delta \hat{\rho}_{-\mathbf{q}},$$

where $d_{\eta s a, \mathbf{k}}$ is the electron annihilation operator and $\varepsilon_a^{(\eta s)}(\mathbf{k})$ is the dispersion of the magnetic subband indexed by $a$. The projected density operator subtracting a background charge is given by:

$$\delta \hat{\rho}_{\mathbf{q}} = \sum_{\eta s, ab, \mathbf{k}} \langle \Psi_a^{(\eta s)}(\mathbf{k}) | e^{-i \mathbf{q} \cdot \mathbf{r}} | \Psi_b^{(\eta s)}(\mathbf{k} + \mathbf{q}) \rangle \left( d^\dagger_{\eta s a, \mathbf{k}} d_{\eta s b, \mathbf{k}+\mathbf{q}} - \frac{1}{2} \delta_{\mathbf{q}, \mathbf{G}} \delta_{a,b} \right),$$

where $\mathbf{G} = m \mathbf{g}_1 + n \mathbf{g}_2, \{m, n\} \in Z$. In the literature the coefficient due to projection is referred to as the structure factor $\hat{\Lambda}_{\mathbf{q}}^{(\eta s)}(\mathbf{k})$, and can be equivalently denoted as:

$$\left[ \hat{\Lambda}_{\mathbf{q}}^{(\eta s)}(\mathbf{k}) \right]_{a,b} \equiv \langle \Psi_a^{(\eta s)}(\mathbf{k}) | e^{-i \mathbf{q} \cdot \mathbf{r}} | \Psi_b^{(\eta s)}(\mathbf{k} + \mathbf{q}) \rangle \equiv \langle u_a^{(\eta s)}(\mathbf{k}) | u_b^{(\eta s)}(\mathbf{k} + \mathbf{q}) \rangle,$$

where $|u_a^{(\eta s)}(\mathbf{k})\rangle$ is the magnetic unit cell periodic part of the magnetic Bloch state. $A$ is the total area of the system, and $V_{\mathbf{q}} = \frac{2\pi e^2}{\epsilon_0 \epsilon_r} \frac{\tanh(|\mathbf{q}|\xi/2)}{|\mathbf{q}|}$ is the dual gate screened Coulomb interaction.

The parameters for the theoretical calculations in the main text is as follows: twist angle $\theta = 1.05°$, uniaxial heterostrain strength $\epsilon = 0.2\%$ and orientation $\varphi = 0°$, monolayer graphene Fermi velocity $v_F = 9.264 \times 10^5 m/s$, interlayer tunneling parameters $w_0 = 77 meV$ and $w_1 = 110 meV$, relative dielectric constant $\epsilon_r = 15$, and screening length $\xi = 4\sqrt{|L_1||L_2|} \approx 52 nm$. Lattice relaxation effects are neglected in this calculation. For details of the self-consistent Hartree-Fock method we refer readers to SI of Ref. [1]. Most generally, our Hartree-Fock procedure can probe Slater determinant states given by the following equation:

$$|\Omega_{\text{HF}}\rangle = \prod_{n, \mathbf{k}}' \left( \sum_{sa} \alpha_{sa, \mathbf{k}}^{(n)} d^\dagger_{Ksa, \mathbf{k}} + \sum_{s'a'} \beta_{s'a', \mathbf{k}+\mathbf{q}_0}^{(n)} d^\dagger_{K's'a', \mathbf{k}+\mathbf{q}_0} \right) |0\rangle,$$

where $\prod_{n, \mathbf{k}}'$ is a constrained product over all occupied single-electron eigenstates $\{n, \mathbf{k}\}$ of the Hartree-Fock Hamiltonian. $\mathbf{q}_0$ is an arbitrary wavevector shift between single electron states in opposite valleys, and $\{\alpha_{sa, \mathbf{k}}^{(n)}, \beta_{s'a', \mathbf{k}+\mathbf{q}_0}^{(n)}\}$ are variational parameters satisfying $\sum_{sa} |\alpha_{sa, \mathbf{k}}^{(n)}|^2 + \sum_{s'a'} |\beta_{s'a', \mathbf{k}+\mathbf{q}_0}^{(n)}|^2 = 1$ for any $\{n, \mathbf{k}\}$. The Hartree-Fock Hamiltonian is minimized with respect to both $\{\alpha_{sa, \mathbf{k}}^{(n)}, \beta_{s'a', \mathbf{k}+\mathbf{q}_0}^{(n)}\}$ and $\mathbf{q}_0$. The latter gives the variational freedom to probe intervalley Kekulé spiral ordered (IKS) states as well as other translation symmetry breaking states such as the symmetry-broken Chern insulator (SBCI). We hereby define the one-particle density matrix:

$$[\hat{P}(\mathbf{k})]_{\eta s a, \eta' s' a'} = \langle d^\dagger_{\eta s a, \mathbf{k}} d_{\eta' s' a', \mathbf{k}+\mathbf{q}_0} \rangle.$$

Its diagonal matrix elements denote the occupation number of a given electronic state. For an IKS state [4] the density matrix also satisfies the condition:

$$\hat{t}_{L_2}[\hat{P}(k)]_{\eta s a, \eta' s' a'} \hat{t}_{L_2}^{-1} = e^{i\varphi(\eta-\eta')}\left[\hat{P}\left(k + \frac{\phi}{\phi_0}g_1\right)\right]_{\eta s a, \eta' s' a'}.$$

where an IKS state constrains $\varphi$ to be an arbitrary real number, see SI III. D of Ref [1] for details.

### STM Local Density of States (LDOS)

Here we present a calculation of the LDOS in finite magnetic field shown in Fig. 3 of the main text and Fig. S11 of the Supplementary Information, which would be relevant for STM experiments. We denote the eigenbasis of the Hartree-Fock mean field Hamiltonian and its single particle eigenenergies as: $\{|\Gamma_{n,k}\rangle, E_{n,k}\}$. The LDOS at a given energy $\mu$ is therefore given by [5]:

$$\mathcal{N}(\mu, r) \propto \sum_{n,k} \delta(\mu - E_{n,k})|\langle r|\Gamma_{n,k}\rangle|^2.$$

Here the LDOS does not resolve the sublattice and layers. The real space wavefunction is related to the magnetic Bloch states via a unitary transformation:

$$\langle r|\Gamma_{n,k}\rangle = \sum_{\eta s a} U_{\eta s a, n}(k)\langle r|\Psi_a^{(\eta s)}(k)\rangle,$$

and the real space wavefunction of the magnetic Bloch states can be calculated from the monolayer graphene Landau level basis states (see SI of Ref. [1]). In computing the LDOS maps for SBCIs in the main text, we approximated $\delta(\mu - E_{n,k}) \approx \frac{1}{\pi}\frac{\gamma}{(\mu-E_{n,k})^2+\gamma^2}$, with an energy broadening factor $\gamma = 0.5 meV$. In Fig. 3 and Fig. S11, the LDOS are plotted using an effective periodicity of the strained moiré superlattice, $a_M \equiv \sqrt{|L_1||L_2|}$.

## Note S2. Quantum geometry properties of the FQHs and the FCI indicators.

The experimentally observed fractional quantum Hall states (FQHs) emanate from the charge neutrality point (CNP), and importantly require a critical field of around 4 Tesla to occur. The latter condition places the observations in the Hofstadter regime when the magnetic length is comparable to the moiré length scale, in sharp contrast to FQHs emanating from a single Landau level. The dissimilarity between the (interaction-renormalized) magnetic subbands and single LLs is demonstrated by computing the Berry curvature and quantum geometry variations in the magnetic Brillouin zone, as plotted in Fig. 4 of the main text and Fig. S15 of the Supplementary Information, with the theory shown here.

For a group of $N$ magnetic subbands with a finite Chern number $C \neq 0$ and separated from the rest of the spectrum by an energy gap, we define the multiband quantum geometric tensor:

$$\eta_{\mu\nu}^{mn}(k) = N\mathcal{A}\langle\partial_\mu u_{m,k}|[1-P(k)]|\partial_\nu u_{n,k}\rangle,$$

where $\partial_\mu \equiv \partial_{k_\mu}$ is the derivative with respect to the magnetic wavevector $k$, $\langle r|u_{n,k}\rangle \equiv \langle r|e^{-ik\cdot r}|\Gamma_{n,k}\rangle$ is the periodic part of the mean-field eigenstates, $P(k) = \sum_{n=1}^{N}|u_{n,k}\rangle\langle u_{n,k}|$ is the projector onto this group of magnetic subbands, and $\mathcal{A}$ is the area of the magnetic Brillouin zone. The Berry connection and quantum geometric tensor can also be derived from the appropriate action of the position operator on the magnetic translation group irreps [6].

The symmetric and antisymmetric part of the quantum geometric tensor,

$$g_{\mu\nu}^{mn} = \frac{1}{2}(\eta_{\mu\nu}^{mn} + \eta_{\nu\mu}^{mn}), \mathcal{F}^{ab} = i(\eta_{xy}^{mn} - \eta_{yx}^{mn}),$$

are used to formulate the ideal quantum geometry indicators [7]. Considering the gate-screened short range interaction, this is related to the possibilities of hosting fractionalized states; while longer range interactions can change this. Specifically, we define:

$$\sigma[\mathcal{F}] \equiv Tr\left[\left(\frac{1}{2\pi C}\mathcal{F} - 1\right)^2\right]^{\frac{1}{2}} \text{ where } Tr[O] \equiv \frac{1}{N\mathcal{A}}\sum_{n=1}^{N}\int d^2k O_{nn}(k).$$

The deviation of $\sigma[\mathcal{F}]$ from 0 provides a measure of the Berry curvature uniformity in the magnetic Brillouin zone. We also define the trace condition:

$$T[\eta] \equiv Tr(g) - |Tr(\mathcal{F})| \geq 0, g \equiv g_{xx} + g_{yy}.$$

Deviation of $T[\eta]$ from 0 provides a measure of how different the wavefunctions are to those of the lowest Landau level (LLL).

Numerically, $\mathcal{F}(k)$ can be computed by forming an infinitesimal closed loop near $k$:

$$\mathcal{F}(k) \approx -Im\log\left[\hat{\Lambda}_{q_x}(k)\hat{\Lambda}_{q_y}(k+q_x)\hat{\Lambda}_{-q_x}(k+q_x+q_y)\hat{\Lambda}_{-q_y}(k+q_y)\right].$$

Here $[\hat{\Lambda}_q(k)]_{mn} = \langle u_{n,k}|u_{m,k+q}\rangle$ is the structure factor matrix defined with respect to the group of (interaction-renormalized) subbands of interest, and not to be confused with the non-interacting structure factor defined in previous section. This procedure is inherently gauge-invariant. Similarly, $g(k)$ can be computed by noting that:

$$[\hat{\Lambda}_q(k)\hat{\Lambda}_{-q}(k+q)]_{mn} \approx \delta_{mn} - \sum_{\mu\nu}q_\mu q_\nu g_{\mu\nu}^{mn}(k)$$

whose trace is gauge-invariant. In Fig. 4 of the main text and Fig. S15 of the Supplementary Information, we plot the Berry curvature and quantum metric in the magnetic Brillouin zone for the relevant magnetic subband hosting the experimentally observed FQH. In these plots, $q_M$ is defined following $q_M \equiv \sqrt{|g_1||g_2|}/q$ at flux ratio $\Phi/\Phi_0 = p/q$.

A comparison with calculations done for the non-interacting magnetic subband near CNP (Fig. S16) shows that the strong Coulomb interaction improves the uniformity of the Berry curvature

(and quantum geometry) of the magnetic subbands even in the presence of substantial heterostrain.

## Note S3. Orbital decomposition of the magnetic subbands

An alternative way of interpreting the FQHs is to make analogies to the single LL physics of a monolayer graphene. We can express the magnetic subband wavefunctions in terms of the monolayer LL wavefunctions. Specifically for the Landau gauge used here, we denote the monolayer LL wavefunction as $\left|\psi_{Nl,r}^{(\eta s)}(k_2)\right\rangle$, where $l = t, b$ is the layer index and $N = 0, \pm 1, \pm 2, \ldots$ is the LL index. In above notation we replaced the quantum number $k_y \in (-\infty, \infty)$ by:

$$k_y = \frac{2\pi}{|L_2|}\left(k_2 + r\frac{1}{q}\right), k_2 \in [0,1/q), r \in \mathbb{Z},$$

yielding a consistent definition of $k_2$ as previously where it is used to denote the quantum number within the first magnetic Brillouin zone. We refer interested readers to Ref. [1] SI Sec. II for a detailed derivation. The single-electron eigenstates $|\Gamma_{n,k}\rangle$ of the Hartree-Fock Hamiltonian can be expressed as:

$$|\Gamma_{n,k}\rangle = \sum_{\eta s, Nlr} \widetilde{U}_{\eta sNlr,n}(k) \left|\psi_{Nl,r}^{(\eta s)}\left(k_2 + r\frac{1}{q}\right)\right\rangle.$$

where $\widetilde{U}_{\eta sNlr,n}(k)$ is a unitary matrix at every $k$ from the Hartree-Fock calculations. It relates the single-electron eigenstates to the monolayer LL wavefunctions. The orbital decomposition of the magnetic subband ($n_0$) relevant for the experimentally observed FQHs can therefore be calculated as:

$$W_N^{(n_0)} = \sum_{\eta s, lr, k} \left|\widetilde{U}_{\eta sNlr,n}(k)\right|^2.$$

If the FQH is born out of single LL physics, then we expect the orbital decomposition to be strongly peaked at a single LL index N. In Figure S14, we show that the relevant magnetic subband mixes many LLs, and therefore inconsistent with FQH emerging from a single LL.

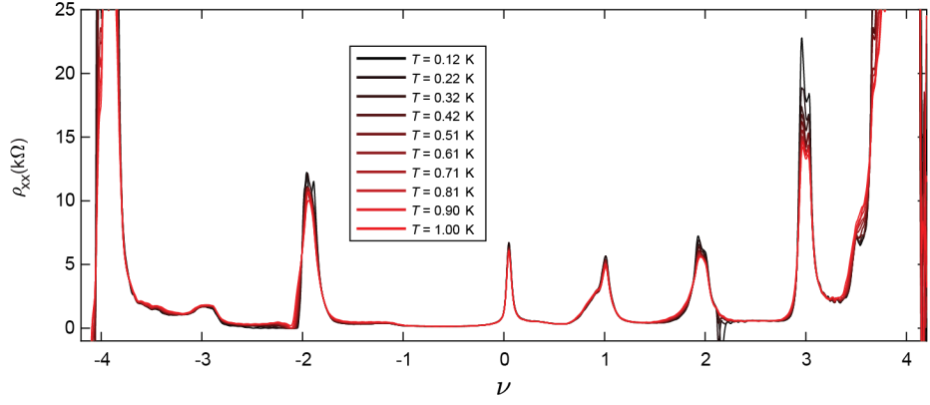

**Figure S1. Temperature dependence of the resistivity $\rho_{xx}$ at zero magnetic field.** At integer filling factors within the flat bands, we observe weakly insulating states at $\nu$ = -2, 2, 3. The resistive peak at $\nu$ = -3, 0, 1 is nearly temperature independent below 1 K. No resistive peak is observed for $\nu$ = -1.

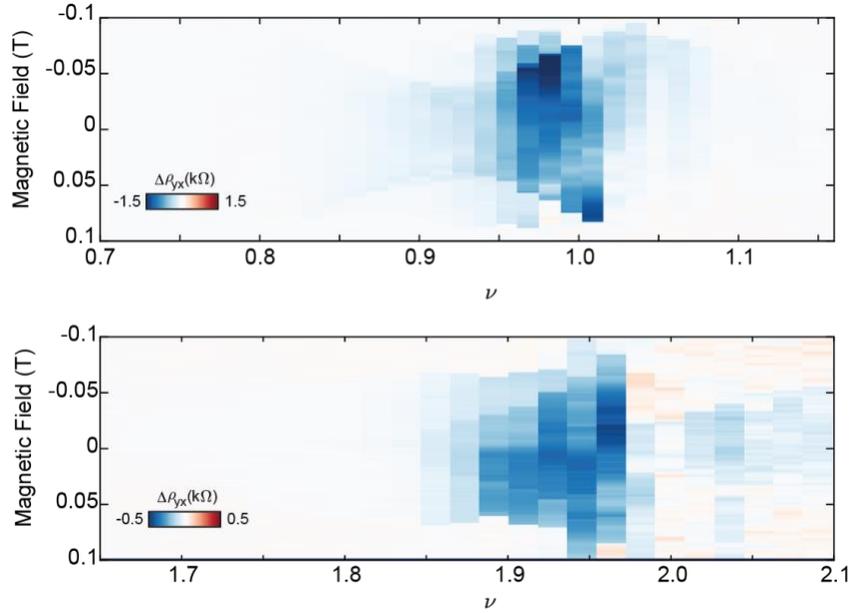

**Figure S2. Filling factor dependence of AHE near $\nu$ = 1 and 2.** The $\nu$ dependence of the AHE near $\nu$ = 1 and 2 are shown in top and bottom panels respectively. We plot the amplitude of the hysteresis loop in Hall resistivity, $\Delta\rho_{yx} = \rho_{yx}^{B\downarrow} - \rho_{yx}^{B\uparrow}$.

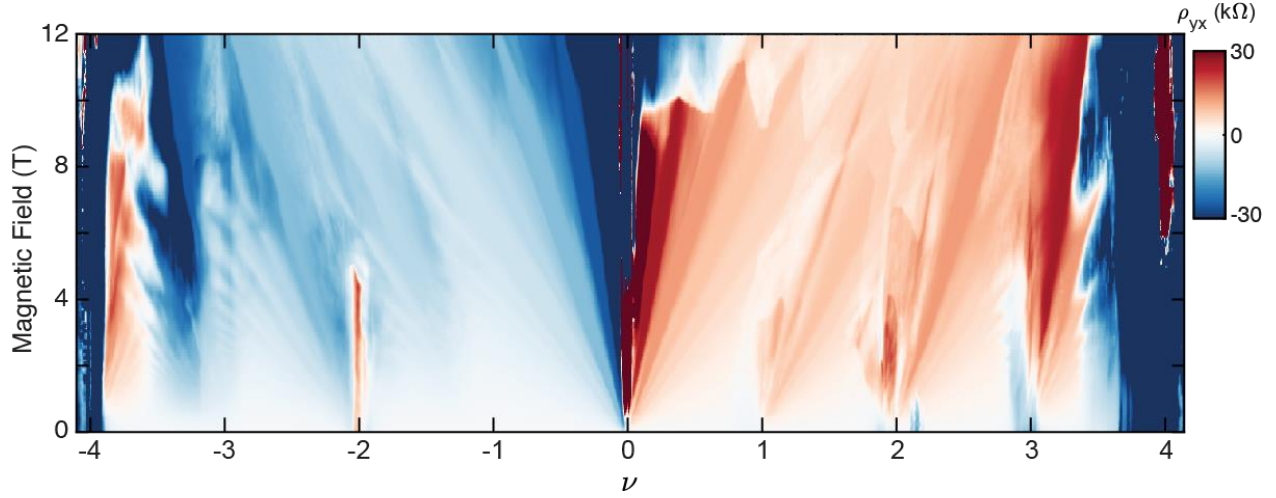

**Figure S3. Landau Fan diagram in Hall resistivity $\rho_{yx}$ at $T = 340$ mK.** For measurements at $\nu > 0$ ($\nu < 0$), the graphene contacts are slightly electron (hole) doped to achieve optimal quality.

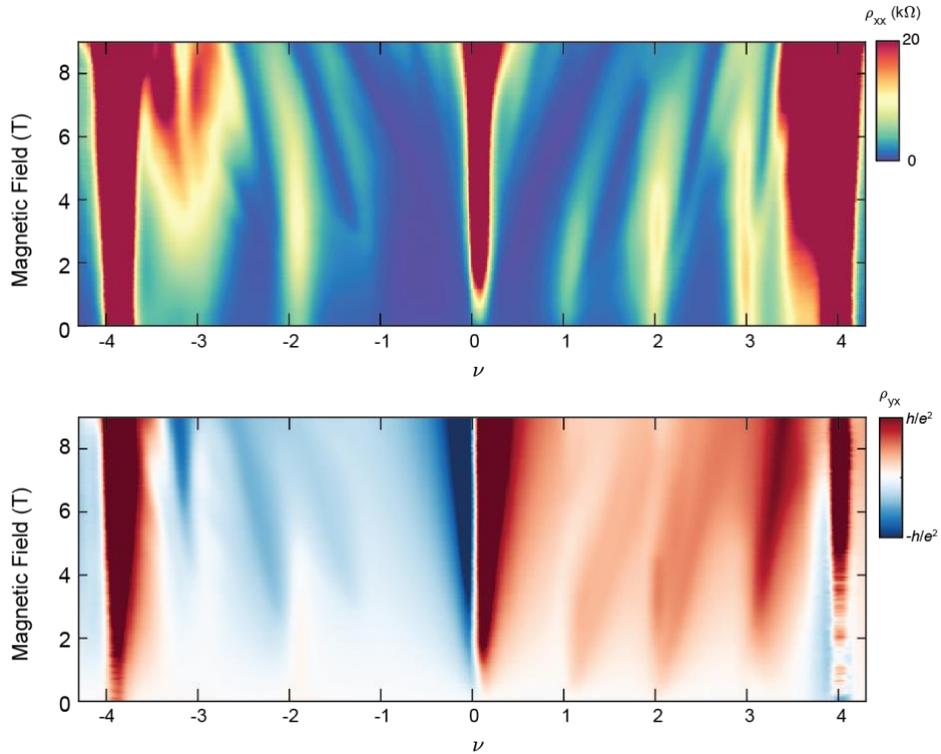

**Figure S4. Landau Fan diagrams in $\rho_{xx}$ and $\rho_{yx}$ at $T = 2$ K.** At high temperature, correlated states with larger energy gaps survive. In particular, we note the observation of the main sequence correlated Chern insulators (CCI) with total Chern number $|t| = 1, 2, 3$ and moiré filling factor $|s| = 3, 2, 1$, as well as the integer quantum Hall (IQH) state with $|t| = 4$.

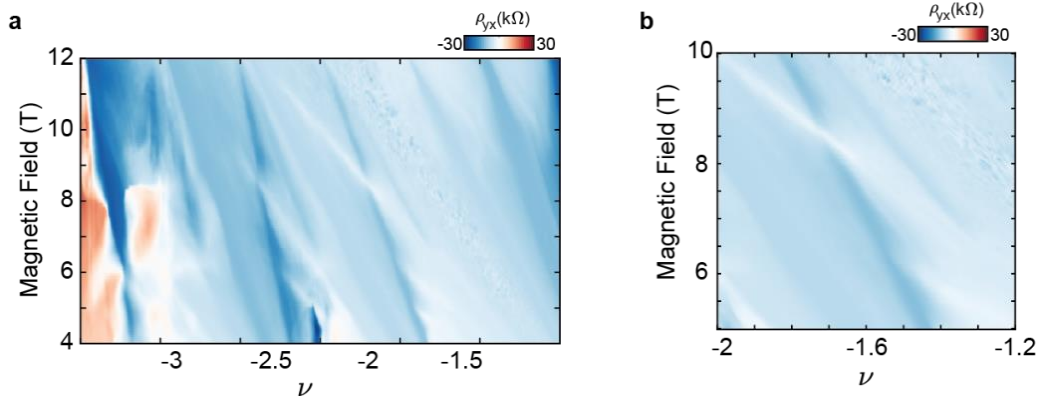

**Figure S5. Landau Fan diagrams in $\rho_{yx}$ of the SBCIs.** Zoom-in measurement of the $\rho_{yx}$ Landau fan focusing on the SBCI states on the hole-doped side. **a** and **b** panels show $\rho_{yx}$ in the same phase space as shown in Fig. 2a and c.

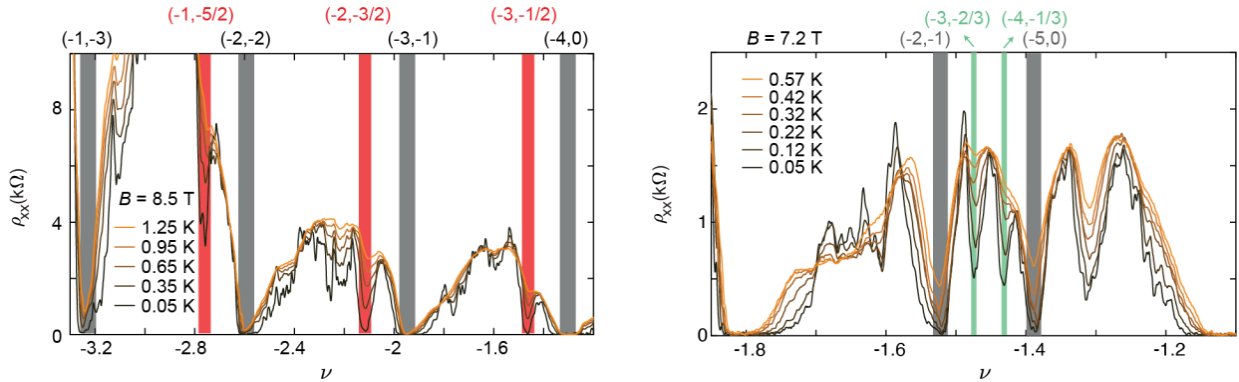

**Figure S6. Temperature dependence of the SBCIs.** The suppressed resistivity $\rho_{xx}$ at low temperature show the clear formation of the SBCIs, marked by the red and green shaded regions.

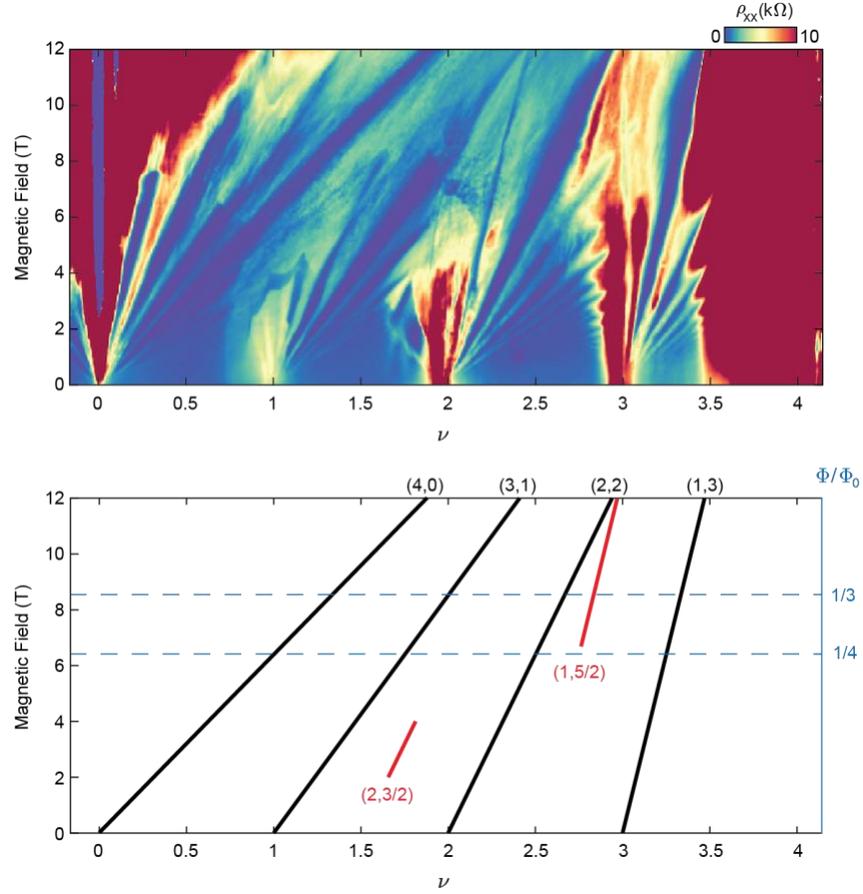

**Figure S7. Observation of SBCIs at electron doping side.** Landau fan diagram in resistivity $\rho_{xx}$ measured at $T = 300$ mK, focusing on the phase space $\nu > 0$. We note the observation of (1, 5/2) SBCI state, with the same onset magnetic field/magnetic flux ratio $\Phi/\Phi_0 = 1/4$. More interestingly, the (2, 3/2) SBCI only shows in a narrow range of magnetic field below a quarter $\Phi/\Phi_0$.

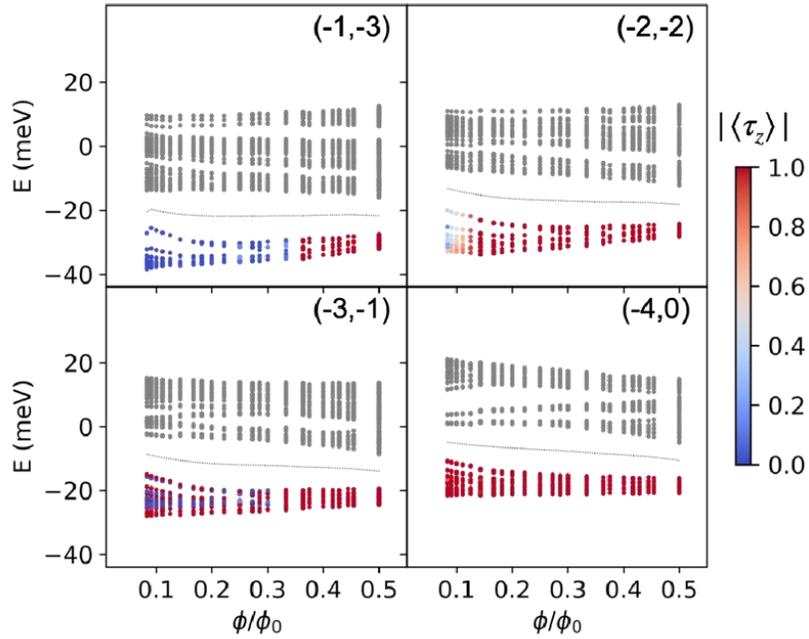

**Figure S8. Flavor symmetry of the main sequence Correlated Chern insulators with $(t, s) =$ (-1, -3), (-2, -2), (-3, -1).** The valley and spin symmetric IQH along (-4,0) is plotted for reference. The occupied states are colored by their valley polarization $\langle \tau_z \rangle$. $\langle \tau_z \rangle \to 1$ means maximal valley polarization whereas $\langle \tau_z \rangle \to 0$ means maximal intervalley mixing. For (-1, -3), (-2, -2), (-3, -1), the CCIs experience a phase transition from a correlated Hofstadter ferromagnet (CHF) at high magnetic field to an IKS state at low magnetic field. Figure reproduced from Ref. [1] SI.

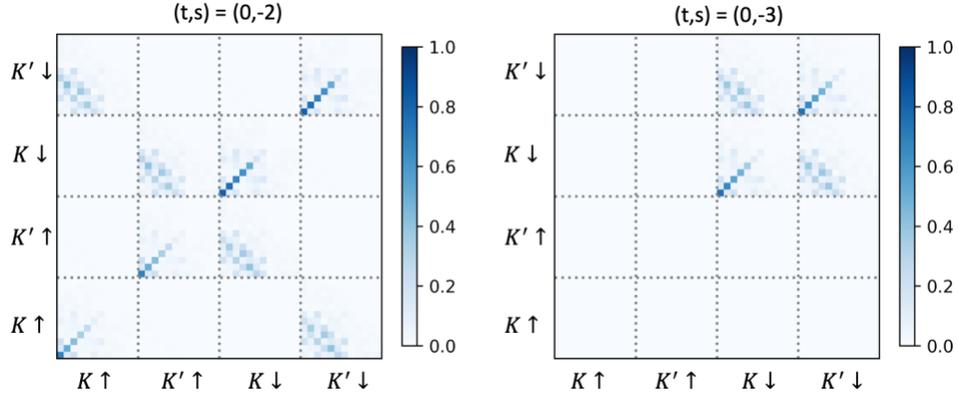

**Figure S9. Flavor symmetry of the $t = 0$ IKS states at finite magnetic field.** Representative density matrices (absolute value) at $\Gamma$ point of the magnetic Brillouin zone for: left panel, (0,-2); right panel, (0,-3) IKS states respectively. Calculation is done at $\Phi/\Phi_0 = 1/6$. For the heterostrain we used in the calculation, the IKS wavevector is $\boldsymbol{Q}_{IKS} = \boldsymbol{g}_1/2$. We caution however that the $\boldsymbol{Q}_{IKS}$ can change subject to uniaxial heterostrain condition in relevant experiments.

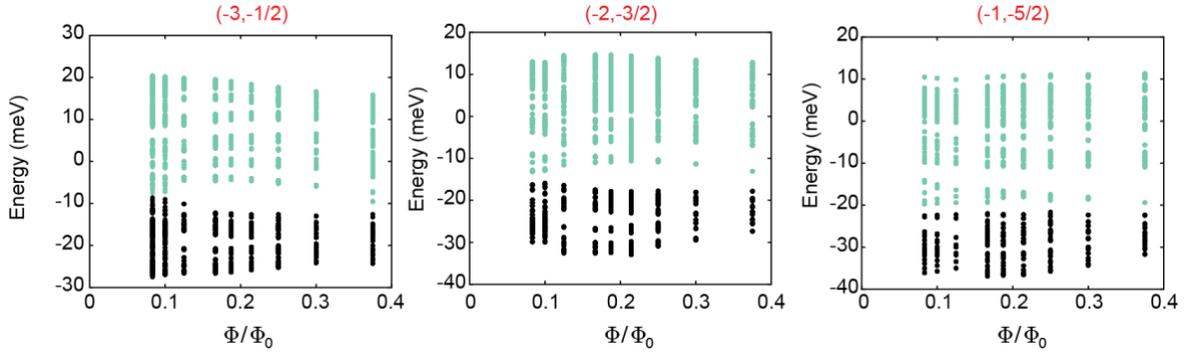

**Figure S10. Interacting Hofstadter spectrum of the half-integer SBCIs.** Energy spectra vs flux are calculated when the Fermi energy is within the gap of the three half-integer SBCIs. We use the same color scheme as in Fig. 3 to denote the occupied (green) and unoccupied (black) states.

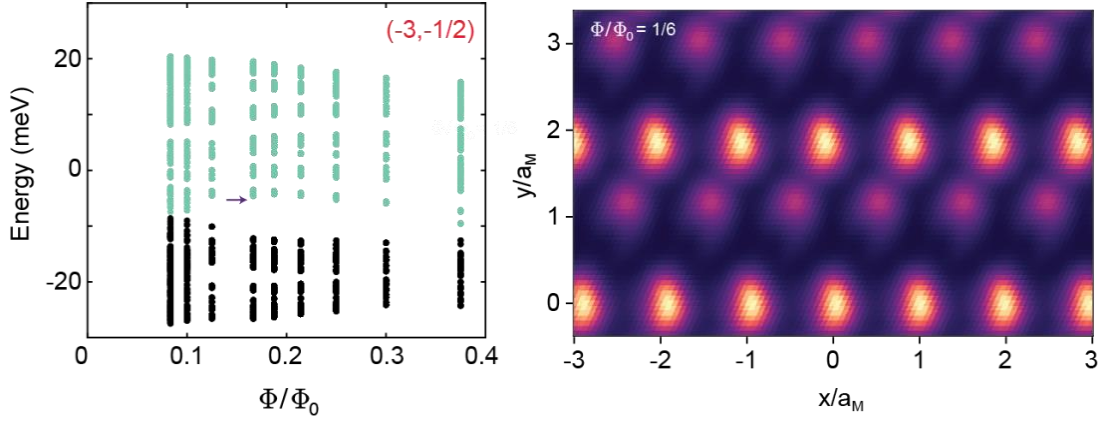

**Figure S11. Local density of states at the conduction band edge of the (-3, -1/2) SBCI.** Calculation is done at $\Phi/\Phi_0 = 1/6$, pointed out by the purple arrow in the left energy spectrum. $a_M$ ($a_M \equiv \sqrt{|L_1||L_2|}$) is the effective periodicity of the strained moiré superlattice. The stripe-like charge density distribution breaks the moiré translation symmetry, similar to that at the valence band edge of the SBCI in Fig. 3 of the main text.

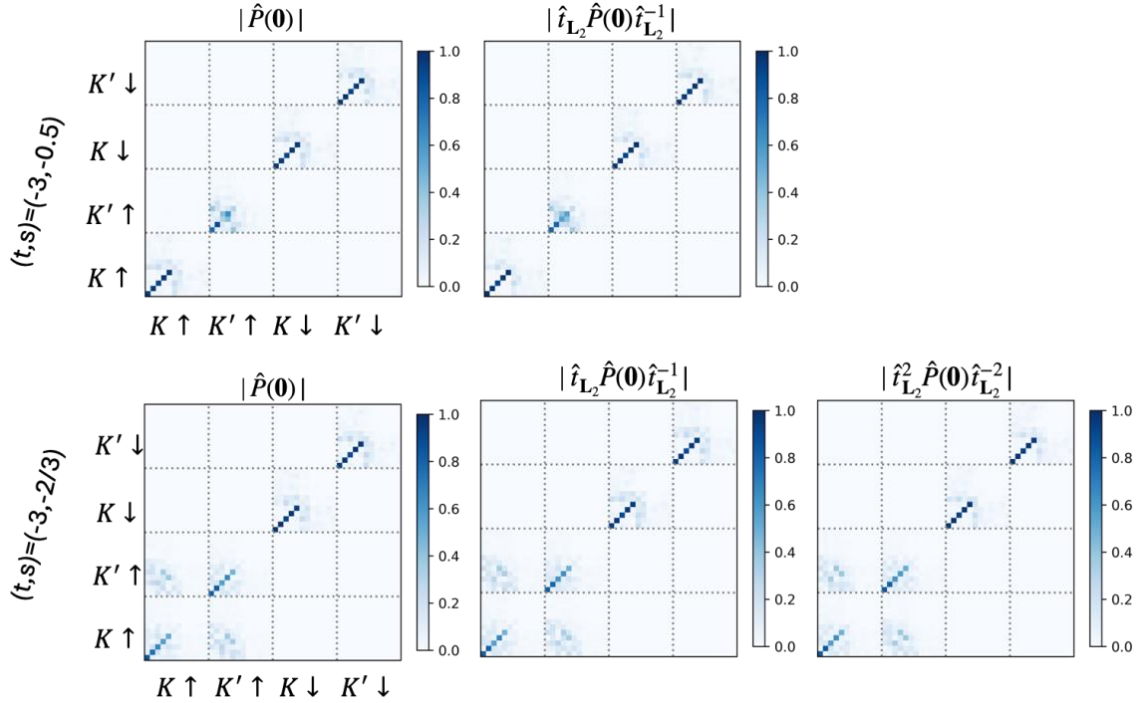

**Figure S12. Flavor and magnetic translation symmetry of SBCI states at finite magnetic field.** Representative density matrices (absolute value) at a few magnetic wavevectors for: upper panel, $(t, s) = (-3, -1/2)$; lower panel, $(-3, -2/3)$ SBCI states respectively. Calculation is done at $\Phi/\Phi_0 = 1/6$. The period of the stripe is identified via $\hat{t}_{L_2}$ translations. For $(-3, -1/2)$, the density matrix repeats upon $\hat{t}_{L_2}^2$, corresponding to a period - 2 stripe. For $(-3,-2/3)$ it is a period - 3 stripe.

The SBCI states can be either valley and spin polarized (upper panel) or developing intervalley coherence (lower panel). However, due to the small Hartree-Fock energy differences (~ 0.05 meV per moiré unit cell), we do not postulate if either type of SBCIs will be observed in experiments.

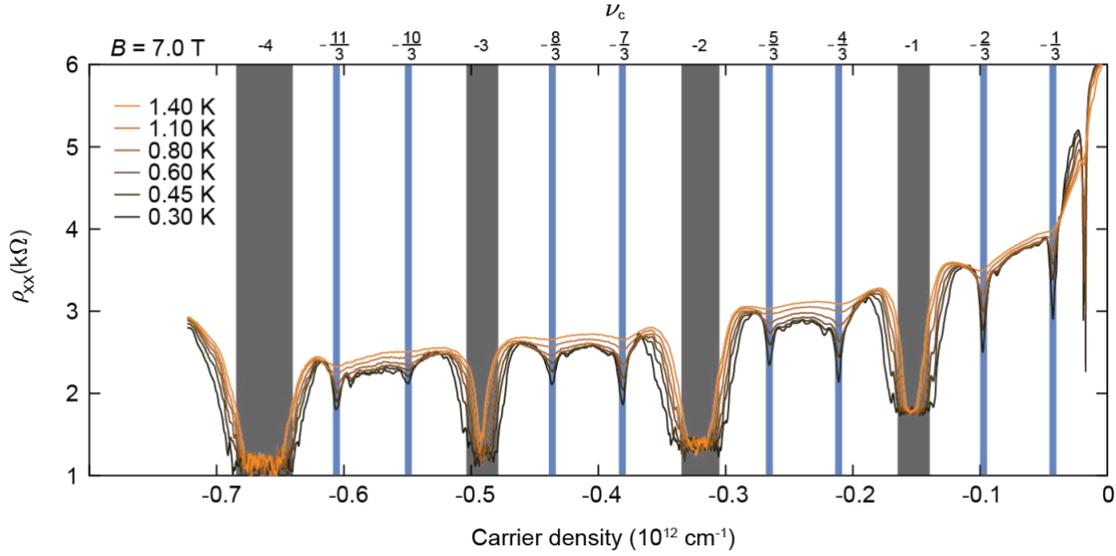

**Figure S13. Temperature dependence of the FQHs.** Resistivity $\rho_{xx}$ of the FQHs measured at $B = 7$ T and various temperature. The FQHs with denominator 3 is marked by the blue shaded region, where we obtain the thermal activation gap as shown in Fig. 4b inset.

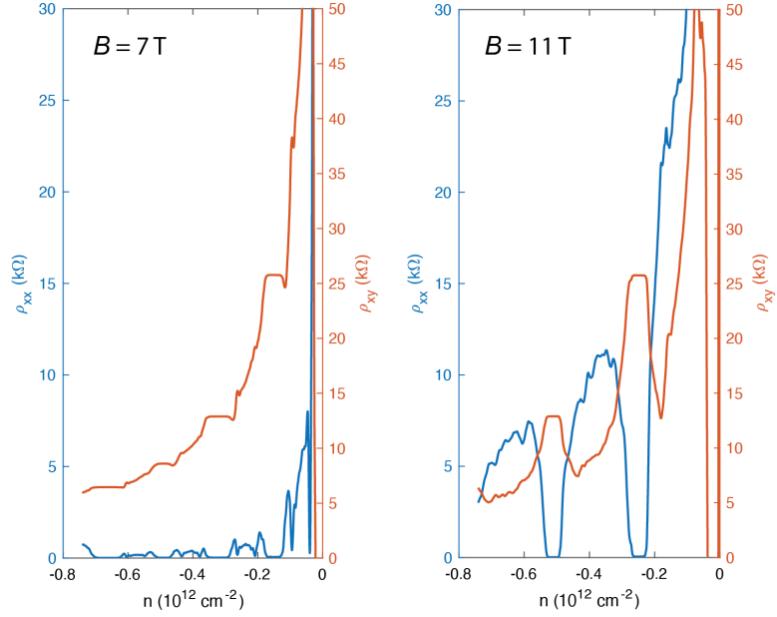

**Figure S14. FQHs and Fermi liquid at high magnetic field.** Resistivity $\rho_{xx}$ and Hall resistivity $\rho_{xy}$ of the FQHs measured at $B = 7$ T and 11 T. At high magnetic field, the FQH states disappear and transition into a dissipative Fermi liquid phase at partial fillings of the magnetic subbands.

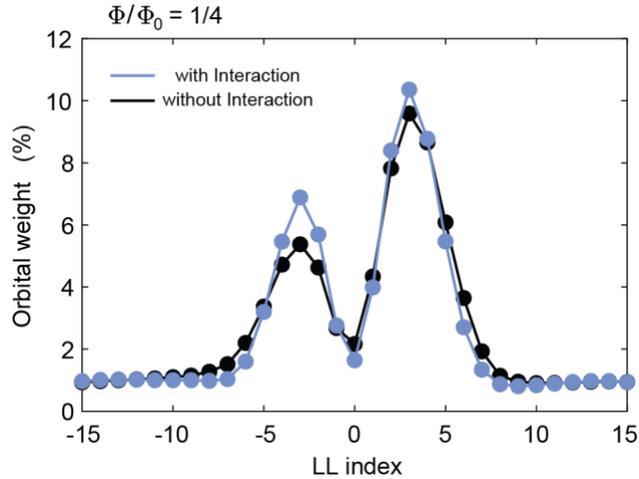

**Figure S15. Orbital decomposition of the magnetic subband.** Decomposition of the FQH's parent magnetic subband, in both the interacting Hofstadter spectrum (blue) and the non-interacting single-particle Hofstadter spectrum (black). In both cases, the magnetic subband has strong orbital weight at high $N$ LLs peaked at $N = 3$ and $-3$. The Coulomb interaction reduces the number of LL mixings, pushing the subband wavefunctions closer to that of a higher $N$ LL.

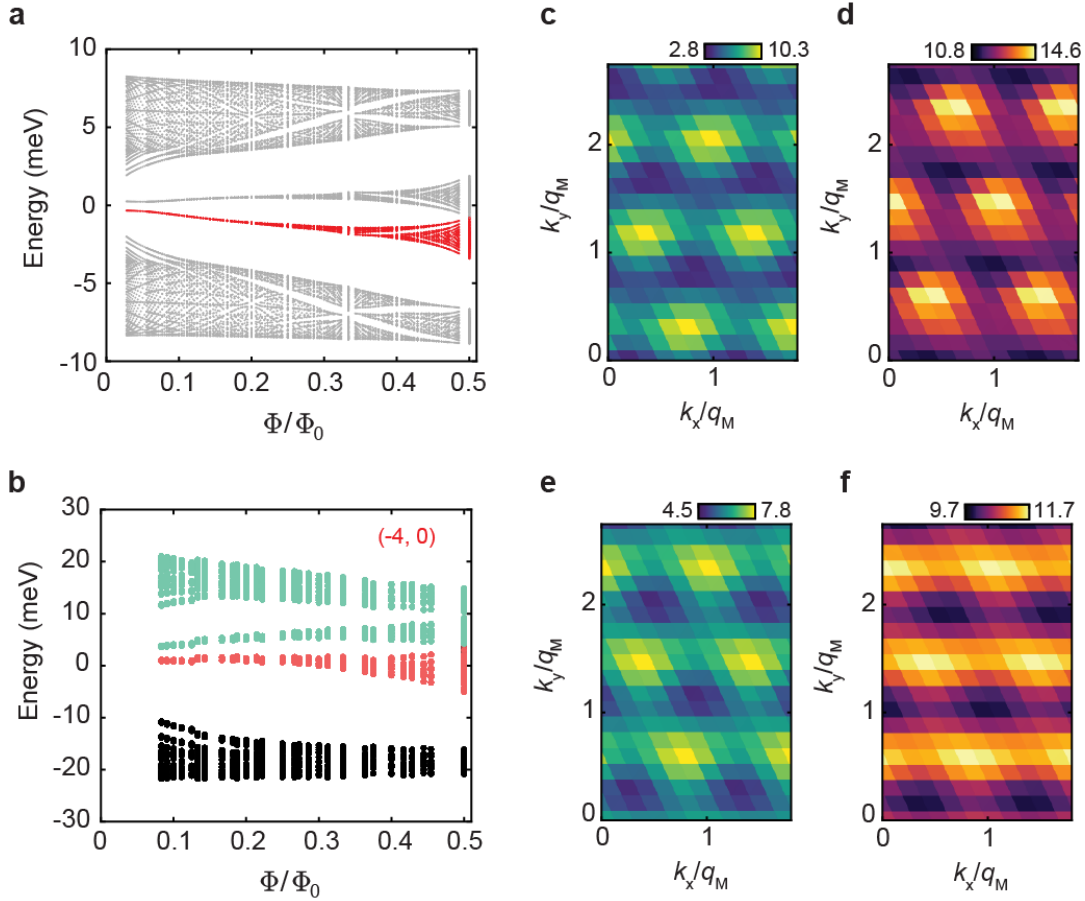

**Figure S16. Comparison of the quantum geometry conditions with and without Coulomb interaction. a-b,** Non-interacting Hofstadter spectrum and interacting Hofstadter spectrum (Fermi energy within (-4,0) IQH gap) respectively, both plotted for the K valley with spin down. The parent subbands of the FQHs are marked red. The Coulomb interaction significantly broaden the bandwidth of that from a single-particle calculation. **c-d,** Berry curvature $\mathcal{F}$ and quantum metric $g$ of the non-interacting subband respectively, calculated at $\Phi/\Phi_0 = 1/4$. **e-f,** Berry curvature $\mathcal{F}$ and quantum metric $g$ of the interacting subbands respectively, calculated at $\Phi/\Phi_0 = 1/4$. $q_M$ is defined following $q_M \equiv \sqrt{|g_1||g_2|}/q$ at flux ratio $\Phi/\Phi_0 = p/q$. Absolute values of $\mathcal{F}$ and $g$ are plotted instead of normalized values. Except for the spatial shift of the $\mathcal{F}$ and $g$ hotspot, we highlight that the spatial distribution of the interacting subband is much smoother compared with the non-interacting subband.